# A discrete model and analysis of one-dimensional deformations in a structural interface with micro-rotations

Aleksey A. Vasiliev [a], Andrey E. Miroshnichenko [b,*], Massimo Ruzzene [c]

[a] *Department of Mathematical Modelling, Tver State University, 35 Sadoviy per., 170002 Tver, Russia*
[b] *Nonlinear Physics Centre, Research School of Physical Sciences and Engineering,*
*The Australian National University, Canberra ACT 0200, Australia*
[c] *School of Aerospace Engineering, Georgia Institute of Technology,*
*270 Ferst Drive, Atlanta GA, 30332-0150, United States*

**Abstract**

The static and dynamic properties of a Cosserat-type lattice interface of finite thickness are studied, so that both displacements and rotational degrees of freedom are taken into account. The model allows considering interfaces with a beam-like microstructure and interfaces with finite size particles as particular cases. One-dimensional solutions describing shear and micro-rotations at the interface are obtained and discussed. Harmonic as well as localized solutions, and the properties of the interfaces as filters for elastic waves are investigated. It is shown that both systems with long- and systems with short-wavelength localization may exist.

## 1. Introduction

The problem of modeling and studying the properties of structural interfaces is one of the most urgent theoretical and practical problems in mechanics. In many cases, the hypothesis of zero-thickness or the absence of interfaces can be used. However, the interface's finite thickness and structure often needs to be considered (Bigoni and Movchan, 2002) for the adequate description of interface properties and their influence on phenomena such as, among others, fracture, plasticity, and dissipation. This conceptual idea is used in the modeling of different systems and at different scale levels. Structural interfaces appear in nature, for example, in biological systems (Bigoni and Movchan, 2002; Bertoldi et al., 2007a), where the development of an adequate interface theory is particularly important. The analysis of the interface between an inclusion and a matrix as a structural interface with finite thickness is also of relevance for problems in mechanics of composites (Bertoldi et al., 2007b). Similarly, it may sometimes be important to consider the surface of a body as a layer of finite thickness with an internal microstructure for the adequate modeling of surface effects, such as stress concentrations and surface waves localized in the narrow boundary zone. One can sometimes consider brittle, slip, and shear zones as having zero thickness, but it is often necessary to consider them as finite thickness structural interfaces. In addition, in structural engineering applications, sandwich beams and plates often play the role of structural interfaces.

* Corresponding author. Tel.: +61 2 6125 9653; fax: +61 2 6125 8588.
*E-mail address:* andrey.miroshnichenko@anu.edu.au (A.E. Miroshnichenko).

One of the attractive features of structural interfaces with periodic structure is their ability to serve as filters for elastic waves over certain frequency bands (Bigoni and Movchan, 2002). Similar filtering capabilities have also been documented for sandwich beams with regular and re-entrant (auxetic) honeycomb cores, for honeycomb and auxetic lattices and periodic composite materials (Hussein et al., 2002; Ruzzene and Scarpa 2003, 2005; Phani et al., 2006).

We show that the changes of the displacements and rotations of elements in narrow zones, which appear near boundaries, concentrated forces, defects or inhomogeneous regions in diamond type lattice interfaces, may have a monotone or a short-wavelength localized character. Lattice truss structures with diamond cell, and the elastic properties, buckling and crushing of truss-core plates were investigated, for example, in Zupan et al. (2004), Queheillalt and Wadley (2005). The comparison of generalized continuum and discrete modeling of cellular structures, and the calculation of the stress concentration factor around a circular notch in 2-D cellular materials, in particular with diamond cell structure, were considered in Kumar and McDowell (2004).

## 2. Discrete model

We consider a square Cosserat-type lattice interface (Figs. 1 and 3), whose behavior is described by the two displacement degrees of freedom $u_n$, $v_n$, the rotational degree of freedom $\varphi_n$. The potential energy associated with the elastic connection of elements $k$, $m$ can be written in local coordinates in the following form

$$E_{pot}^{k,m} = \frac{1}{2} K_n^{k,m} (u_m - u_k)^2 + \frac{1}{2} K_s^{k,m} \left[ v_m - v_k - r_{k,m} \frac{\varphi_m + \varphi_k}{2} \right]^2 + \frac{1}{2} G_r^{k,m} (\varphi_m - \varphi_k)^2, \qquad (1)$$

where $r_{k,m}$ is a length parameter, while $K_n^{k,m}$, $K_s^{k,m}$ and $G_r^{k,m}$, respectively define the elastic resistance to relative longitudinal and transverse displacements, and rotations of neighbouring elements.

The representation of the potential energy in the form (1) is similar to that which is assumed in micropolar theory elasticity (Eringen, 1968) and also used in discrete models of granular media (Limat, 1988; Suiker et al., 2001) and in models of micro- and nano-scale thin films (Randow et al., 2006). The potential energy of beam elements (Fig. 2a) used for modeling of the beam lattices and for the construction of continuum models for bodies with beam-like microstructure (Noor, 1988), is a particular case of the model (1) and may be obtained by letting in Eq. (1) $K_n = AE/h$, $K_s = 12EI/h^3$, $G_r = EI/h$, $r_{k,m} = h$, where $A$, $E$, $I$, $h$ are the cross-sectional area, the Young's modulus, the second moment of area of the beam cross section and the length of the beam.

The potential energy in the form (1) may be also used to model the interaction of finite size particles (Fig. 2b), as considered in Pavlov et al. (2006) and Potapov et al. (2009), where the values of microstructural parameters for some crystals were obtained from experimental data. The potential energy of the elastic bonds is $E = (k/2)[(L_1 - L_0)/L_0]^2$, where $k$ is the bond stiffness, $L_0$ and $L_1$ are the bond lengths before and after deformation, respectively. We linearize the change in the bond length, $\Delta L = L_1 - L_0$, with respect to displacements and rotations assuming that they are small (Lisina et al., 2001; Vasiliev et al., 2005). The potential energy of the system is equal to the sum of the potential ener-gies of the bonds. By comparison with Eq. (1) one obtains the following expressions $K_n = k_0/h^2 + 2k_1/l^2 + 2k_2 l^2/l_d^4$, $K_s = 2k_2(2a)^2/l_d^4$, $G_r = 2k_1 a^2/l^2$, $r_{k,m} = h$, where $k_0$, $k_1$, $k_2$ are stiffnesses of the bonds with the lengths $h$, $l = h - 2a$ and $l_d = \sqrt{l^2 + (2a)^2}$, respectively (Fig. 2b).

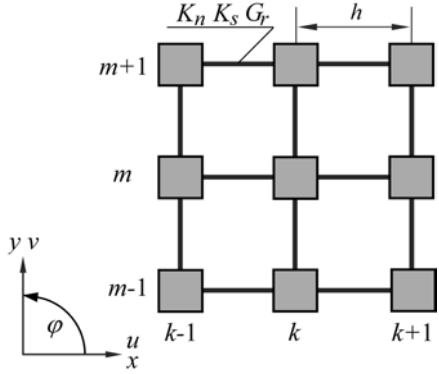 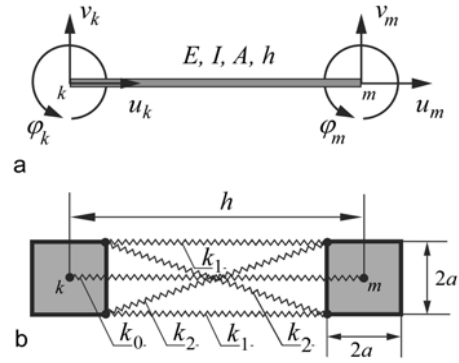

Fig. 1. Square Cosserat lattice. Coordinate system and considered notations.

Fig. 2. Possible realizations of the Cosserat lattice model with potential energy of the form (1).

The kinetic energy of the elements is described in the form $E_{kin}^n = \frac{1}{2} M \dot{u}_n^2 + \frac{1}{2} M \dot{v}_n^2 + \frac{1}{2} J \dot{\varphi}_n^2$, where $M$ is the mass and $J$ is the moment of inertia of the $n$th element. The equations of motion are obtained by using Lagrange's equations.

We consider one-dimensional shear deformations of a structural interface of finite thickness between two rigid bodies (see Fig. 3). The problem of tension at the interface, comparing discrete and different generalized continuum models were considered in Vasiliev et al. (2008). Assuming that the generalized displacements are constant for elements along the diagonals, i.e. for elements with $k + m = const$, we denote components of displacement and rotation in a new coordinate system $O\xi\eta$ by using the abbreviated notation $U_m$, $V_m$, and $\Phi_m$. The equations for $U_m$ and $V_m$, $\Phi_m$ are decoupled. The problem for $U_m$ was considered in Vasiliev et al. (2008). We will consider the equations for $V_m$ and $\Phi_m$ only

$$M\ddot{V}_m = K_n(V_{m+1} - 2V_m + V_{m-1}) + K_s[V_{m+1} - 2V_m + V_{m-1} - d(\Phi_{m+1} - \Phi_{m-1})],$$
$$J\ddot{\Phi}_m = (2G_r - K_s d^2)(\Phi_{m+1} - 2\Phi_m + \Phi_{m-1}) + K_s d(V_{m+1} - V_{m-1} - 4d\Phi_m), \quad (2)$$

where $d = \sqrt{2} h/2$ is distance between layers in the $O\xi$ direction.

## 3. Static one-dimensional solutions

By substituting $V_m = Ce^{\lambda m}$, $\Phi_m = \alpha Ce^{\lambda m}$ into Eq. (2) we obtain the characteristic equations

$$e^{2\lambda} - 2e^\lambda + 1 = 0, \qquad (3)$$
$$e^{2\lambda} - 2\gamma e^\lambda + 1 = 0, \qquad (4)$$

where $\gamma = (\delta + 1)/(\delta - 1)$, $\delta = 2(K_n + K_s)G_r / K_n K_s d^2$.

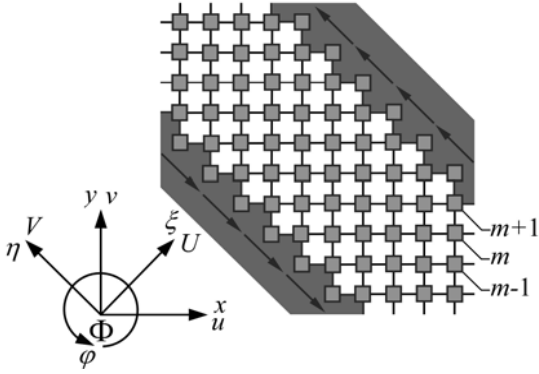 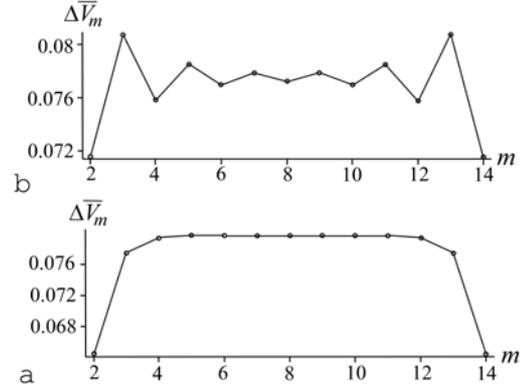

Fig. 3. Structural lattice interface of finite size in different coordinate systems.

Fig. 4. The differences of the displacements of neighbouring elements $\Delta \overline{V}_m = \overline{V}_m - \overline{V}_{m-1}$ for parameters: (a) $\overline{K}_n = 2$, $\overline{G}_r = 0.03$ and (b) $\overline{K}_n = 2$, $\overline{G}_r = 0.6$. Continuous lines are drawn to underline short- and long- wavelength behaviour in boundary regions.

Equation (3) has a solution $\lambda = 0$ of the second order which gives the linear part of the static solution. The second part of the solution corresponds to Eq. (4). One can show that $\gamma^2 - 1 > 0$ for non-negative values of the constants $K_n > 0$, $K_s \geq 0$, $G_r \geq 0$. Moreover, taking into account Vieta's theorem, we can conclude that $\exp(\lambda_+)\exp(\lambda_-) = 1$, $\exp(\lambda_+) + \exp(\lambda_-) = 2\gamma$. Hence, the solutions are real $\lambda_\pm = \ln\left[\gamma \pm \sqrt{\gamma^2 - 1}\right]$ in the case $\gamma > 0$, i.e. for $\delta > 1$, and complex $\lambda_\pm = \ln\left[|\gamma| \mp \sqrt{\gamma^2 - 1}\right] + i\pi$ for $\gamma < 0$, i.e. for $0 < \delta < 1$. So the second part of the static solution has an exponentially decaying character for parameters $2(K_n + K_s)G_r - d^2 K_n K_s > 0$. In this case the general solution has the form

$$V_m = C_1 + C_2 m + C_3 e^{\lambda m} + C_4 e^{-\lambda m},$$
$$\Phi_m = \frac{1}{2d}C_2 + \frac{K_n + K_s}{K_s d}\frac{\sinh(\lambda/2)}{\cosh(\lambda/2)}\left[e^{\lambda m}C_3 - e^{-\lambda m}C_4\right]. \quad (5)$$

Short wavelength spatially localized deformations take place for parameters $2(K_n + K_s)G_r - d^2 K_n K_s < 0$. The general solution for such parameters has the form

$$V_m = C_1 + C_2 m + (-1)^m e^{\lambda m} C_3 + (-1)^m e^{-\lambda m} C_4,$$
$$\Phi_m = \frac{1}{2d}C_2 + \frac{K_n + K_s}{K_s d}\frac{\cosh(\lambda/2)}{\sinh(\lambda/2)}\left[(-1)^m e^{\lambda m}C_3 - (-1)^m e^{-\lambda m}C_4\right]. \quad (6)$$

The parameter $\lambda = \ln\left[|\gamma| + \sqrt{\gamma^2 - 1}\right] > 0$ in the solutions (5) and (6) defines the degree of spatial localization of the decaying solutions.

It is important to note, that there exist systems featuring long-wavelength localization and systems with short-wavelength localization. In order to demonstrate this, we will use the system presented in Fig. 2b. One can consider stiffnesses $k_n$, $n = \overline{0, 2}$, as independent. This allows us to construct systems with different values of the constants $K_n > 0$, $K_s \geq 0$, $G_r \geq 0$ in the phenomenological potential (1) by choosing elastic constants, $k_0 = h^2(4a^2 K_n - 4G_r - l^2 K_s)/4a^2$, $k_1 = G_r l^2/2a^2$, $k_2 = K_s l_d^4/8a^2$, and geometric parameters, $a$ and $h$. The long-wavelength, $\gamma = 1.02 > 0$, solution with the parameter localization $\lambda = 0.2$ occurs, for example, in the system with parameters $k_0 = 0$, $k_2 = 0.01 k_1$, $h = 2\sqrt{2} a$. In the limit, where the parameter $G_r$ is equal to zero, we have the short-wavelength solution, $\gamma = -1 < 0$, without localization, $\lambda = 0$. Accordingly, short-wavelength localized distortions take place in systems with small enough value of $G_r$, which corresponds to zero or relatively small values of the stiffness $k_1$. For example, in the case $k_0 = 0$, $k_1 = 0.01 k_2$, $h = 6\sqrt{2} a$ we have $\gamma = -1.0038$, $\lambda = 0.088$.

For illustration we consider an interface containing 14 layers, $m = \overline{1, 14}$, with boundary conditions $V_1 = -V_*$, $\Phi_1 = 0$, $V_{14} = V_*$, $\Phi_{14} = 0$. Without loss of generality, one can consider three parameters $M$, $d$, and $K_s$ as arbitrary and introduce the dimensionless quantities $\overline{K}_n = K_n/K_s$, $\overline{G}_r = G_r/K_s d^2$, $\overline{V}_m = V_m/V_*$. In order to underline the type of localization we evaluate the difference of the displacements of neighbouring elements $\Delta \overline{V}_m = \overline{V}_m - \overline{V}_{m-1}$. The localization has a rapidly varying character for parameters $\overline{K}_n = 2$, $\overline{G}_r = 0.03$ (see in Fig. 4a) and monotone for $\overline{K}_n = 2$, $\overline{G}_r = 0.6$ (see Fig. 4b). These results indicate that accurate modeling of localized solutions may be very important for the correct description of deformations at interfaces.

## 4. Parametric analysis of dynamic and static solutions

We consider dynamic solutions of the form

$$V_m(t) = \overline{V} e^{i\omega t + Km}, \quad \Phi_m(t) = \overline{\Phi} e^{i\omega t + Km}. \tag{7}$$

Substitution into the discrete equations of motion (2) gives

$$\begin{bmatrix} a_1 + M\omega^2 & -a_2 \\ a_2 & a_3 + J\omega^2 \end{bmatrix} \begin{bmatrix} \overline{V} \\ \overline{\Phi} \end{bmatrix} = \begin{bmatrix} 0 \\ 0 \end{bmatrix}, \tag{8}$$

where $a_1 = 2(K_n + K_s)(\cosh K - 1)$, $a_2 = 2K_s d \sinh K$, $a_3 = 2(2G_r - K_s d^2)(\cosh K - 1) - 4K_s d^2$.

For the analysis of both harmonic and spatially localized excitations we consider solutions (7) with complex values $K = K_{\text{Re}} + iK_{\text{Im}}$. Nontrivial solutions of the system (8) exist when its determinant is equal to zero

$$MJ\omega^4 + (Ja_1 + Ma_3)\omega^2 + a_1 a_3 + a_2^2 = 0. \tag{9}$$

This condition defines the dispersion curves $\omega = \omega(K_{\text{Im}}, K_{\text{Re}})$ in the space $K_{\text{Im}} K_{\text{Re}} \omega$.

Letting $K = iK_{\text{Im}}$ and introducing the new variables $\Omega = \omega^2$, $Z = 4\sin^2(K_{\text{Im}}/2)$ gives the following form for the dispersion relations (9)

$$b_{11} Z^2 + 2b_{12} Z\Omega + b_{22} \Omega^2 + b_1 Z + b_2 \Omega = 0, \tag{10}$$

where $b_{11} = 2(K_n + K_s)G_r - K_n K_s d^2$, $2b_{12} = (K_s d^2 - 2G_r)M - (K_n + K_s)J$, $b_{22} = MJ$, $b_1 = 4K_n K_s d^2$, $b_2 = -4K_s d^2 M$.

The analysis of dispersion relation in the form (10) in the two-dimensional space $-\infty < Z < \infty$, $\Omega \geq 0$ is useful for the classification and analysis of the dispersions curves defined by Eq. (9) in the three dimensional space $\omega \geq 0$, $K_{Re} \geq 0$, $0 \leq K_{Im} \leq \pi$. Namely, the dispersion curves $\omega = \omega(K_{Im}, K_{Re})$ in the planes $K_{Im} = 0$, $K_{Re} = 0$, and $K_{Im} = \pi$ correspond to parts of the curve $\Omega = \Omega(Z)$ in the intervals $Z < 0$, $0 < Z < 4$, and $Z > 4$, respectively. The branch $\omega = \omega(K_{Im}, K_{Re})$ defined in the area of complex values ($K_{Im} \neq 0$, $K_{Re} \neq 0$) is located at frequencies $\Omega$, at which there are no points of the curve $\Omega = \Omega(Z)$. Examples of the $\Omega(Z)$ curves in dimensionless form $\overline{\Omega} = \Omega/\sqrt{K_s/M}$ that are typical for areas with the points 1-7 are shown in Fig. 5. Dimensionless parameter $\overline{J} = J/Md^2$ is chosen equal to $1/8$. In order to illustrate correspondence of $\omega = \omega(K_{Im}, K_{Re})$ and $\Omega = \Omega(Z)$ curves we show additionally dispersion curves for points 1, 2, and 6 in Figs. 6a, 6b, and 6c respectively.

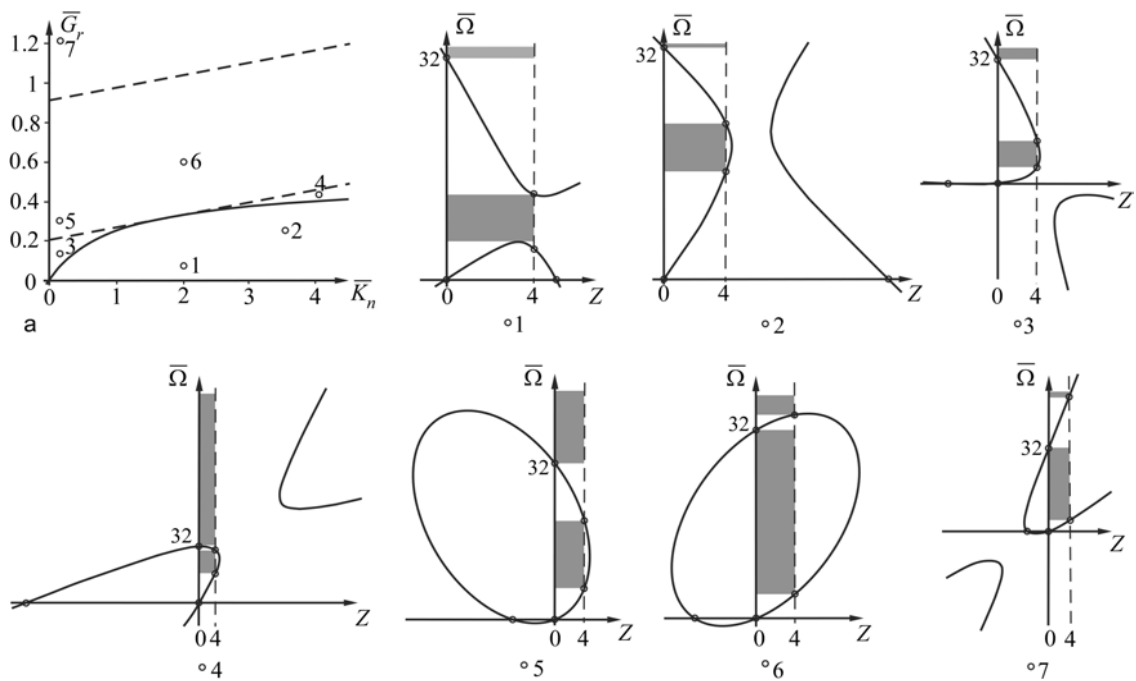

Fig. 5. The area of dimensionless elastic parameters. Some examples of the curves for areas with points 1-7.

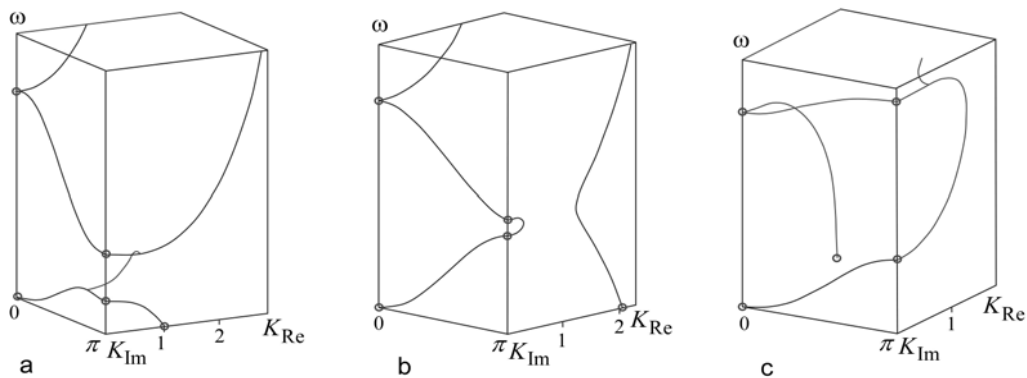

Fig. 6. Dispersion curves $\omega = \omega(K_{Im}, K_{Re})$ for parameters corresponding to points 1, 2 and 6 in Fig. 5a are shown in Figs. a, b, and c respectively.

The type of the curve expressed by Eq. (10) is defined by the sign of $D = b_{11}b_{22} - b_{12}^2$. The curves are elliptical, parabolic or hyperbolic for the cases $D > 0$, $D = 0$ or $D < 0$, respectively. In the case under consideration, the curve is elliptical for parameters in the region

$$K_n J + K_s \left(d\sqrt{M} - \sqrt{J}\right)^2 < 2G_r M < K_n J + K_s \left(d\sqrt{M} + \sqrt{J}\right)^2. \tag{11}$$

Dashed lines defined by Eq. (11) divide region of elastic parameters in Fig. 5a into regions of parameters, for which the curves in the plane $Z\Omega$ are elliptic or hyperbolic.

The curve intersects the axis $Z = 0$ and line $Z = 4$ at points $\Omega|_{Z=0} = 0$, $\Omega|_{Z=0} = 4K_s d^2 / J$ and $\Omega|_{Z=4} = 4(K_n + K_s)/M$, $\Omega|_{Z=4} = 8G_r / J$, respectively. In Fig. 5, these points are marked on the lines $Z = 0$ and $Z = 4$ by small circles. There always exist two branches of the curve $\Omega(Z)$ joining these points in the interval $0 < Z < 4$. Hence, there always exist two branches of harmonic waves on the plane $K_{Re} = 0$. One of them is called the acoustical branch. Another is the optical branch of micro-rotations.

There always exist values of $\Omega$, for which there are no corresponding points of the curve $\Omega(Z)$ in the interval $0 \leq Z \leq 4$. The solutions (7) for such frequencies have non-zero values of $K_{Re}$. Therefore, for such frequencies elastic waves spatially decrease into the interface. In other words, the interface can be considered as a filter for them. Such bands are called stop bands or band-gaps, which alternate with propagation frequency ranges, called pass bands. The edges of the stop band regions are defined by the values $\Omega|_{Z=0}$, $\Omega|_{Z=4}$ described above and maximum (minimum) values of the curve $\Omega(Z)$ in the cases where they take place inside the interval $0 < Z < 4$. The stop band regions are shaded in Fig. 5.

There always exist branches of the curve $\Omega(Z)$ defined in half-lines $Z < 0$ and $Z > 4$. Hence, there always exist dispersion curves in the planes $K_{Im} = 0$, $K_{Re} \neq 0$ and $K_{Im} = \pi$, $K_{Re} \neq 0$ and corresponding slowly and rapidly varying spatially decreasing dynamical solutions. There exist systems with $\Omega$, for which there are no corresponding points of the curve $\Omega(Z)$. Dispersion curves $\omega = \omega(K_{Im}, K_{Re})$ for such frequencies are defined for complex values of $K$ with $K_{Im} \neq 0$, $K_{Re} \neq 0$. In particular, such branches always exist for systems with curves $\Omega(Z)$ of elliptic form.

Points $Z$ of intersections of the curve $\Omega(Z)$ with the axis $\Omega = 0$ and corresponding points $(K_{Im}, K_{Re})$ of intersection of the dispersion curves $\omega(K_{Im}, K_{Re})$ with the plane $\omega = 0$ define the components of the static solution. The curves $\Omega(Z)$ intersect the axis $\Omega = 0$ at the points $Z = 0$ and $Z = 4/(1-\delta)$, where $\delta$ is defined in Section 2. The value $(K_{Im}, K_{Re}) = (0, 0)$ corresponding to the root $Z = 0$ gives the linear part of the static solutions (5) and (6). Another point of intersection of the curve $\Omega(Z)$ and the axis $\Omega = 0$ is located on the half-line $Z < 0$ for parameters $\delta > 1$ or on the half-line $Z > 4$ in the case where $0 < \delta < 1$. This means that plane $\omega = 0$ is intersected by the dispersion curve $\omega = \omega(K_{Im}, K_{Re})$ at the point $K_{Re} \neq 0$, $K_{Im} = 0$ for $\delta > 1$ or at the point $K_{Re} \neq 0$, $K_{Im} = \pi$ in case where $0 < \delta < 1$. Such points of intersections give the slowly, $K_{Im} = 0$, or rapidly, $K_{Im} = \pi$, varying spatially decreasing, $K_{Re} \neq 0$, part of the solutions. The solid line $\delta = 1$ in the plane of elastic parameters $\overline{K}_n$ and $\overline{G}_r$ in Fig. 5a divides all systems into those with slowly or rapidly varying localized solutions.

# 5. Conclusions

In conclusion, we should note that another problem of interface theory is the validity of an assumption about models of interface structures. Static and dynamic problems for a structural interface are considered in this article by using a micropolar-type discrete model in order to take into account the rotational degree of freedom of micro-structural elements.

Homogenized continuum models are very often used to study materials with microstructure. In particular, such models are used for interface modeling. They help to find analytical solutions by using well-developed mathematical methods of continuum mechanics (Bertoldi et al., 2007a, 2007b) or, in cases where it is impossible, to use effective numerical methods based on artificial discretization with cells, which include several unit cells of the body (Noor, 1988; Kumar and McDowell, 2004). Moreover, the fracture, plasticity and stability theories and corresponding criteria are formulated in most cases in terms of continuum mechanics.

The models should take into account essential structural effects. The analysis, which was made in this article for one-dimensional case, shows that the description of not only long- but also short-wavelength exponentially decreasing displacements and rotations in static and dynamic problems might be important for modeling of interfaces. Classical solid mechanics theory is broadly applicable, but often gives considerable errors in the description of such effects. Generalized models are formulated in order to be used in such cases. Comparison of discrete, classical and generalized continuum models for structural interfaces represent natural extension of this work and will be the topic of future investigations.

# Acknowledgments

The work of A.E.M. was supported by the Australian Research Council. The authors would like to thank Dr. David A. Pawell for careful reading of the manuscript and useful remarks.